\documentclass[aps,prl,twocolumn]{revtex4}

\usepackage{graphicx}
\usepackage{amsmath}
\usepackage{amssymb}
\usepackage{dcolumn}

\begin{document}

\newcommand{\kmod}{k_{\mbox{\scriptsize mod}}}
\newcommand{\dmsk}{\marginpar{$\clubsuit$}}

\title{Spontaneously modulated spin textures in a dipolar
spinor Bose-Einstein condensate}
\author{M. Vengalattore$^{1}$, S. R. Leslie$^1$, J. Guzman$^1$ and D. M. Stamper-Kurn$^{1,2}$}
\affiliation{
    $^1$Department of Physics, University of California, Berkeley CA 94720 \\
    $^2$Materials Sciences Division, Lawrence Berkeley National Laboratory, Berkeley, CA 94720}
\date{\today}

\begin{abstract}
Helical spin textures in a $^{87}$Rb $F=1$ spinor Bose-Einstein
condensate are found to decay spontaneously toward a spatially
modulated structure of spin domains.  This evolution is ascribed to
magnetic dipolar interactions that energetically favor the
short-wavelength domains over the long-wavelength spin
helix.  This is confirmed by eliminating the dipolar
interactions by a sequence of rf pulses and observing a suppression of the formation of
the short-range domains. This study confirms the significance of
magnetic dipole interactions in degenerate $^{87}$Rb $F=1$
spinor gases.
\end{abstract}

\maketitle

In a wide range of materials, the competition between short- and
long-range interactions leads to a rich landscape of spatially
modulated phases arising both in equilibrium and as instabilities in
non-equilibrated systems  \cite{cross,seul}.  In classically ordered
systems such as magnetic thin films \cite{garel} and ferrofluids
\cite{oden02ferrofluid}, short-range ferromagnetic interactions are
commonly frustrated by the long-range, anisotropic magnetic dipolar
interaction, rendering homogeneously magnetized systems
intrinsically unstable to various morphologies of magnetic domains
\cite{dabell}.  Long-range interactions are also key ingredients in
many models of strongly correlated electronic systems \cite{dagotto}
and frustrated quantum magnets \cite{scho04qmbook}. 

In light of their relevance in materials science, strong dipole
interactions have been discussed as an important tool for studies of
many-body physics using quantum gases of atoms and molecules,
offering the means for quantum computation \cite{jaks00},
simulations of quantum magnetism \cite{mich06tool} and the
realization of supersolid or crystalline quantum phases
\cite{goral,buch07polar}. However, in most ultracold atomic gases,
the magnetic dipolar interaction is dwarfed by the contact
($s$-wave) interaction. Hence, experimental efforts to attain
dipolar quantum gases have focused on non-alkali atoms, notably
$^{52}$Cr with its large magnetic moment \cite{laha07ferrofluid},
and on polar molecules \cite{doyl04review}. 

In this Letter, we demonstrate  that magnetic dipole interactions
play a critical role in the behaviour of a quantum degenerate $F=1$
spinor Bose gas of $^{87}$Rb.  In this quantum fluid, $s$-wave
collisions yield both a spin-independent and a spin-dependent
contact interaction \cite{Ho,Ohmi98,sten98spin}, with strengths
proportional to $\bar{a} = (2 a_2 + a_0)/3$ and $\Delta a = (a_2 -
a_0)/3$, respectively, where the scattering length $a_{F}$ describes
collisions between particles of total spin $F$.  In $^{87}$Rb, with
$a_0 \, (a_2) = 5.39 \, (5.31)$ nm, the spin-dependent contact
interaction is far weaker than the spin-independent one; 
nevertheless, it is a critical determinant of the magnetic
properties of degenerate $F=1$ $^{87}$Rb gases
\cite{chan05nphys,kron06tune,sadl06symm}.
 The magnetic dipole interaction strength may be
parameterized similarly  by a length $a_{d} = \mu_0 g_F^2
\mu_B^2 m / (12 \pi \hbar^2)$, where $\mu_0$ is the permeability of
vacuum, $g_F = 1/2$ the gyromagnetic ratio, $\mu_B$ the Bohr
magneton and $m$ the atomic mass \cite{sant03}. Given $a_{d}/\Delta
a = 0.4$, the $F=1$ spinor Bose gas of $^{87}$Rb is an essentially
dipolar quantum fluid \cite{pu}.

In our experiment, the influence of dipolar interactions on the spinor gas is
evidenced by the spontaneous dissolution of deliberately imposed 
long-wavelength helical spin textures, in favor of a finely modulated pattern of spin domains.
 We ascribe the emergence of this modulated phase to the
magnetic dipole energy that disfavors the homogenously magnetized
state and drives the fluid toward short-wavelength spin textures.
 To test this ascription, we re-examine the behavior of spin
helices in condensates in which the dipolar interaction is
eliminated using a rapid sequence of rf pulses.   The suppression of the
modulated phase observed in this case confirms the crucial role
of dipolar interactions.

For this work, spin-polarized $^{87}$Rb condensates of up to $2.3(1)
\times 10^6$ atoms in the $|F=1,m_F=-1\rangle$ hyperfine state and
at a kinetic temperature of $T \simeq 50$ nK were confined in a
single-beam optical dipole trap characterized by trap frequencies
$(\omega_x, \omega_y, \omega_z) = 2 \pi \times (39, 440, 4.2)$
s$^{-1}$. The Thomas-Fermi condensate radius in the $\hat{y}$
(vertical) direction ($r_y = 1.8\, \mu$m) was less than the spin
healing length $\xi_S = (8 \pi \, \Delta a \, n_0)^{-1/2} = 2.4\,
\mu$m where $n_0 = 2.3 \times 10^{14}$ cm$^{-3}$ is the peak density
of the condensate.  This results in a spinor gas that is
effectively two-dimensional with regard to spin dynamics.

\begin{figure*}[t]
\centering
\includegraphics[width=0.75\textwidth]{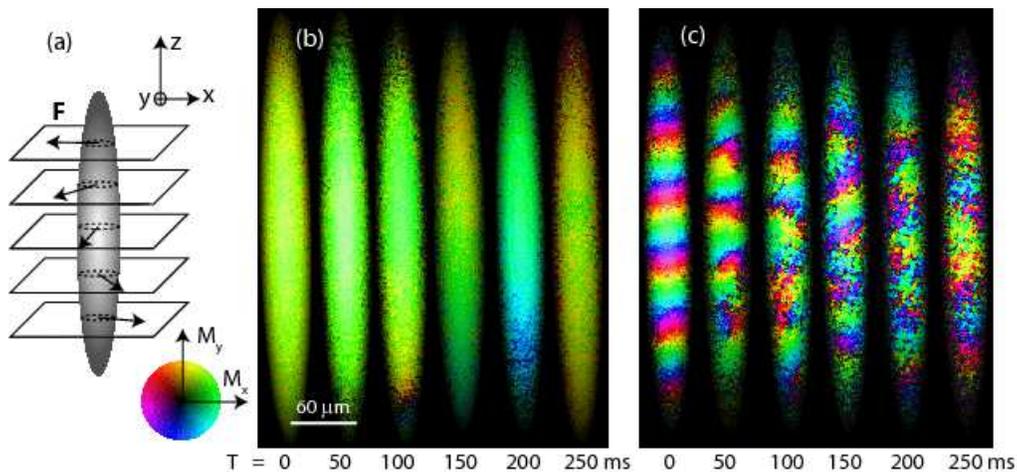}
\caption{Spontaneous dissolution of helical textures in a quantum
degenerate $^{87}$Rb spinor Bose gas.  A transient magnetic field
gradient is used to prepare transversely magnetized (b) uniform or
(a, c) helical magnetization textures. The transverse magnetization
column density after a variable time $T$ of free evolution is shown
in the imaged $x-z$ plane, with orientation indicated by hue and
amplitude by brightness (color wheel shown). (b) A uniform texture
remains homogeneous for long evolution times, while (c) a helical
texture with pitch $\lambda = 60 \, \mu\mbox{m}$ dissolves over
$\sim$200 ms, evolving into a sharply spatially modulated texture.}
\label{fig:dissolutionimage}
\end{figure*}

The condensate was transversely magnetized by applying a $\pi/2$ rf
pulse in the presence of an ambient magnetic field of $B_0 = $
165(5) mG aligned to the $\hat{z}$ axis.  Stray magnetic gradients
(curvatures) were canceled to less than 0.14 mG/cm (4.3 mG/cm$^2$).
A helical spin texture was then prepared by applying a transient
magnetic field gradient $dB_z/dz$ for a period $\tau_p =$ 5 -- 8 ms.
Larmor precession of the atomic spins in this inhomogeneous field
resulted in a spatial spin texture with a local dimensionless spin
of ${\mathbf F} = \cos(\kappa z + \omega_L t)\hat{x} + \sin(\kappa z
+\omega_L t)\hat{y}$, where $\vec{\kappa} = (g_F \mu_B / \hbar)
(dB_z/dz) \tau_p \hat{z}$ is the helix wavevector.  The fast
time variation at the $\omega_L/2 \pi  \simeq 115$ kHz Larmor
precession frequency will be henceforth ignored by considering the
spin at a particular instant in this rapid evolution.  The helix
pitch $\lambda = 2 \pi/\kappa$ ranged between 50 and 150 $\mu$m.
 Given $\lambda \gg \xi_S$, the kinetic energy per atom in this
spin texture, $E_\kappa = \hbar^2 \kappa^2 / 4 m$, was always
negligible compared to the ferromagnetic contact interaction energy
\cite{spinenergyfootnote}. 

The helical spin texture was then allowed to evolve in a homogenous
magnetic field for a variable time  before the vector magnetization
was measured using a sequence of non-destructive phase contrast
images. Because of Larmor precession, a rapid sequence of images
taken with circularly polarized light propagating along the
$\hat{y}$ direction can be analyzed to determine the
column-integrated magnetization perpendicular to the ambient field
\cite{jmhshort,veng07mag}, with vector components $\tilde{M}_{x,y} =
(g_F \mu_B) \tilde{n} F_{x,y}$ where $\tilde{n}$ the column number
density.  Subsequent to this imaging sequence, a $\pi/2$ pulse
was applied to rotate the longitudinal spin $F_z$ into the
transverse spin plane, and a second sequence of images was obtained.
 A least-squares algorithm comparing data from the two imaging
sequences allowed the longitudinal magnetization $\tilde{M}_z$ to be
determined \cite{detailsfootnote}. 

The evolution of helical spin textures is portrayed in Fig.\
\ref{fig:dissolutionimage}. While uniform spin textures ($\lambda
\gg 2 r_z$) remained homogenous for long times, helical textures
($\lambda < 2 r_z$) spontaneously develop short-wavelength
modulations of the magnetization.  This modulated phase is
characterized by spin domains with typical dimensions of $\simeq
10\, \mu$m, much smaller than the pitch of the imprinted helix, with
the magnetization varying sharply between adjacent domains. The
spatial modulations nucleated in regions that varied from shot to shot
but gradually grew to encompass the entire condensate.

\begin{figure}[tb]
\centering
\includegraphics[width=0.37\textwidth]{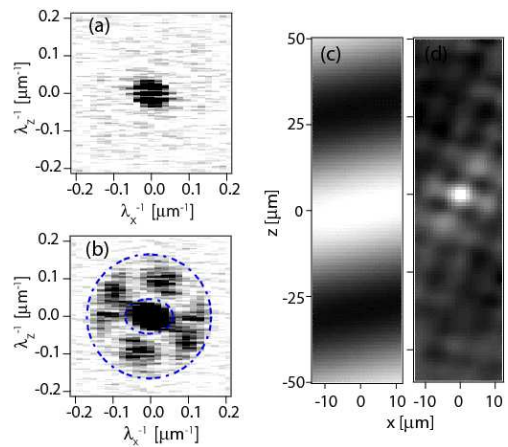}
\caption{Power spectrum of the spatial Fourier transform and the
two-point correlation function $G(x,z)$ for the initial spin helix
(a, c) and the spontaneously modulated phase (b, d). These data are
derived from the same image sequence shown in Fig.\ 1 (d). The
images (a, c) correspond to an evolution time $T = 0$ ms while (b,
d) correspond to an evolution time $T = 250$ ms. The short-range
spatial order is defined as the integrated spectral power in the
annular region shown in (b).} \label{fig:fig2}
\end{figure}

To quantify this behaviour, we considered the power spectrum of the
spatial Fourier transform of the vector magnetization,
$|{\bf{\tilde{M}}}(k_x, k_z)|^2$, where $(k_x, k_z)$ is the spatial
wavevector in the image plane.  This spectrum was found to
consist of two distinct components: a central component that
quantifies the long-range order of the helical texture, and a second
concentration of spectral power at a discrete set of wavevectors of 
magnitude $\kmod \simeq 2 \pi / (10 \, \mu\mbox{m})$ representing the short-range
order of the final modulated texture.  After subtracting out
the background representing image noise, we divided spatial Fourier
space into regions indicated in Fig.\ \ref{fig:fig2} and defined the
integrated spectral power in the central region (annular region) as
the parametrization of long-range (short-range) spatial order in the
quantum fluid. 

\begin{figure}[t]
\centering
\includegraphics[width=0.34\textwidth]{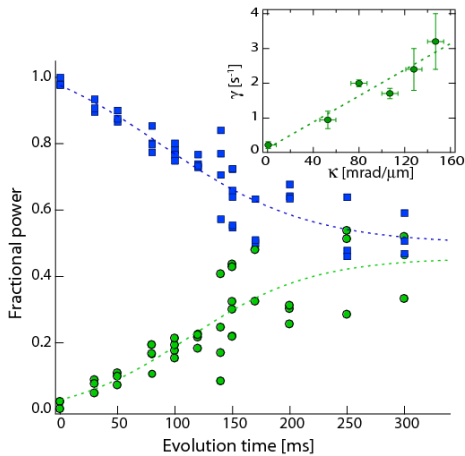}
\caption{Growth of the spontaneously modulated phase ($\bullet$)
coincides with a reduction in the integrated energy in the low
spatial frequency region ($\blacksquare$). The data shown correspond
to an initial helical pitch of 60 $\mu$m. Inset: The initial growth rate
$\gamma$ of the modulated phase as a function of the helix
wavevector. These were extracted from linear fits of the
short-range order parameter at short evolution times.} \label{fig:fig3}
\end{figure}

The formation of the spontaneously modulated texture is reflected in
the reduction of the long-range order parameter and the concomitant
rise of the short-range order parameter (Fig.\ \ref{fig:fig3}).
 During this process, the total spectral power was found to be
roughly constant indicating that the bulk of the quantum fluid
remains fully magnetized even as the long-range order is reduced.
The growth rate $\gamma$ of the short-range order
parameter determined from such data was found to rise monotonically
with the wavevector $\kappa$ of the initial helical texture. While
the long-range order was found to decrease after sufficiently long
evolution times even in condensates prepared with nearly uniform
magnetization, we note that stray magnetic field inhomogeneities of
$\sim$5 $\mu$G across the axial length of the condensate would by
themselves produce a helical winding across the condensate over a
period of 300 ms, constraining our ability to test
the stability of homogenous spin textures.

Another measure of the spontaneous short-range modulation in the
condensate is the appearance of polar-core spin vortices throughout
the gas.  Such vortices were identified as in Ref.\
\cite{sadl06symm} by a net winding of the transverse magnetization
along a closed two-dimensional path of non-zero magnetization in the
imaged gas.  The number of identified spin vortices was roughly
proportional to the short-range order parameter, with no vortices
identified in the initially prepared spin helix and up to 6
vortices/image identified in the strongly modulated texture produced
after free evolution.  In each instance, the number of vortices
with positive and negative charge was found to be approximately
equal.

A striking feature in the evolution of spin textures is the
significant rise in the kinetic energy of the condensed atoms,
reaching a value of $\hbar^2 \kmod^2/ 8 m = h \times 6$ Hz per atom
given that roughly half the spectral weight of the texture's
magnetization is at the wavevector $\kmod$. One expects the total
energy per atom in the condensate to be constant during this
evolution, or even to diminish through the transfer of energy to the
non-condensed portion of the gas. Yet, in examining the energy of
the initially prepared spin helix, we find the local
contact-interaction energy is minimized, the quadratic Zeeman energy
is just $q/2 = h \times 1$ Hz at the ambient magnetic field, and the
kinetic energy of the spin helix is just $E_\kappa/2 < h \times 0.5$
Hz for a helix pitch of $\lambda > 50 \, \mu$m.

This apparent energetic deficit of the spin helix state can be
accounted for by the magnetic dipole interaction.  The on-axis
magnetic field produced by a spin helix in an infinite axial column
of gas with a gaussian transverse density profile can be simply
calculated. From this calculation, we estimate that a gas
with uniform transverse magnetization possesses an excess of $E_d =
\mu_0 g_F^2 \mu_B^2 n_0/2 \, \sim h \times 5$ Hz compared to the
energy of a tightly wound helix, a figure that closely matches the
excess kinetic energy of the finely modulated texture.

To confirm the role of magnetic dipolar interactions in the
evolution of these spin textures, we employed a modification of the
NMR technique of spin-flip narrowing \cite{slic78magres} to
eliminate effectively the dipolar interactions. The interaction energy 
of two magnetic dipoles separated by the displacement vector ${\bf r}$ 
is proportional to ${\bf F}_1 \cdot {\bf F}_2 - 3 (\hat{r} \cdot {\bf{F}}_1) 
(\hat{r}\cdot {\bf{F}}_2)$. If both dipoles experience rapid, common 
rotations that evenly sample the entire $SO(3)$ group of rotations, the 
interaction energy will average to zero regardless of the relative orientations of the
spin vectors ${\bf F}_{1,2}$ and of the displacement vector ${\bf r}$. We note 
that such spin rotations also annul the quadratic Zeeman shift. However, since 
the character of the spontaneously modulated phase was observed to be unchanged 
as $q$ was varied over a factor of 5 ($0.4 < \frac{1}{h}\, q/2 < 2$ Hz), we ignore this small 
difference. 

\begin{figure}[tb]
\centering
\includegraphics[width=0.42\textwidth]{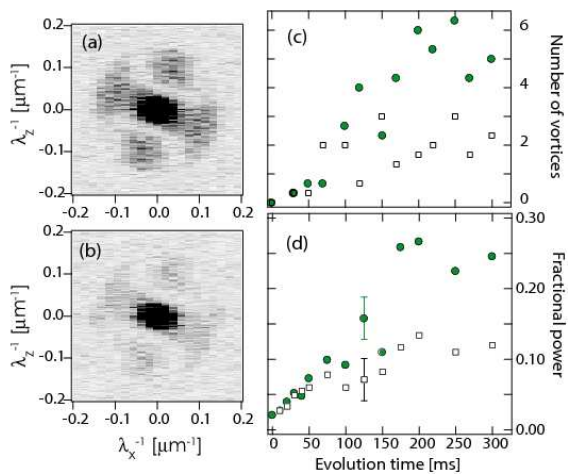}
\caption{Spatial power spectra (a) without or (b) with the
application of rapid rf pulses during the 200 ms evolution following
the preparation of the helical spin texture, are averaged over four
experimental repetitions. Eliminating the dipolar interactions
suppresses the short-range spatial modulation of the final spin
texture.  (c) The average number of detected spin vortices and (d)
the short-range order parameter are shown vs.\ evolution time.
Square symbols indicate data obtained when the dipolar interaction
is spatially averaged.  In all cases, the initial helix pitch was
$\lambda = 80 \, \mu \mbox{m}$.} \label{fig:withandwithout}
\end{figure}

Experimentally, after the initial spin texture had been prepared as
before, we effected such spin rotations by applying a rapid sequence
of $\pi/2$ rf pulses to the Larmor precessing atoms at random
intervals, and, thus, along random rotation axes, at a mean rate of
1 -- 2 kHz.  The spin helix was allowed to evolve under the constant
action of these dipole-cancelation pulses until the pulses ceased
and the sample was imaged as described earlier.  As shown in Fig.\
\ref{fig:withandwithout}, in the absence of dipole interactions a
spin helix was still prone to spontaneous spatial modulation,
consistent with the predictions of dynamic instabilities made in
Refs.\ \cite{lama07helix} and \cite{cher07helix}. However, the depth
of this modulation was significantly suppressed, as seen from the
smaller spectral weight of short-range magnetic order.  This suppression was
less evident for large helix wavevectors ($\kappa > 110$ mrad$/\mu$m), presumably due to the faster growth
of the spontaneously modulated phase. The reduction of the excess kinetic
energy of the final spin texture in the spin-rotation-averaged sample supports
our identification of the dipole energy as its source. 

Finally, we note the distinct six-fold structure in the spatial
Fourier spectrum of the spontaneously modulated phase (Fig.\
\ref{fig:fig2}(b)). The interpretation of this structure is aided by
considering the spatial correlation function of the magnetization,
which we define as
\begin{equation} G(\delta {\bf r}) =
\frac{\sum_{\bf r} \tilde{\bf{M}}({\bf r} + \delta {\bf{r}} ) \cdot
\tilde{\bf{M}}({\bf r})}{(g_F \mu_B)^2 \sum_{\bf r} \tilde{n} ({\bf
r} + \delta {\bf{r}} ) \tilde{n} ({\bf r}) }
\end{equation} 
For the initial spin helix, this correlation function shows the
long-range sinusoidal variation of the transverse magnetization at
the corresponding pitch. In contrast, the final texture shows
significant short-range correlations in both the $\hat{x}$ and
$\hat{z}$ directions, with regions of opposite alignment arranged in
the form of a checkerboard with a lattice spacing $l_m \simeq 10 \,
\mu$m (Fig.\ 2(d)). While these correlations are strongest at short range, they
persist, albeit with diminished strength, even for separations
$\delta r \gg l_m$.   This lattice structure is suppressed in
helices evolving under the active cancelation of dipolar
interactions.

In conclusion, we observe a magnetic-dipole-mediated instability
resulting in the emergence of spontaneously ordered spin domains in
a quantum degenerate spinor Bose gas. The demonstration of the
significance of magnetic dipolar interactions in a $^{87}$Rb spinor
gas presents a new arena for the study of dipolar quantum gases.
 Of particular interest is the influence of such anisotropic
interactions on the true ground state of these spinor gases and the
question of whether the self-organized modulated spin texture
represents an unforeseen equilibrium phase of this dipolar quantum
fluid.

We acknowledge insightful discussions with E.\ Demler, C.\ H.\ Greene and
A.\ Lamacraft.  This work was supported by the NSF, the David and Lucile
Packard Foundation, and DARPA's OLE program. Partial personnel and
equipment support was provided by the Division of Materials Sciences
and Engineering, Office of Basic Energy Sciences. S.\ R.\ L.\
acknowledges support from the NSERC.


\begin{references}
\bibitem{detailsfootnote} Further details on the imaging system and signal analysis will
  be discussed elsewhere.
\end{references}

\end{document}